\documentclass{PoS}
\usepackage{color}
\let\normalcolor\relax
\usepackage{amsmath, amsfonts, bbm, dsfont, mathrsfs,amssymb}
\usepackage{dcolumn}
\usepackage{psfrag}
\usepackage{epsfig}

\newcolumntype{C}{>{$}c<{$}}
\newcolumntype{R}{>{$}r<{$}}
\newcolumntype{L}{>{$}l<{$}}
\newcommand{\SU}{{\rm SU}}
\newcommand{\sci}[1]{\cdot10^{#1}}
\newcommand{\xiav}{\langle\,\xi\,\rangle}
\newcommand{\xisqav}{\langle\,\xi^2\,\rangle}

\title{2+1 flavour Domain Wall Fermion simulations by the RBC and UKQCD collaborations}

\ShortTitle{2+1 flavour DWF simulations}

\author{\speaker{Peter Boyle}\thanks{University of Edinburgh.}\\
        RBC and UKQCD collaborations\\
        E-mail: \email{paboyle@ph-ppt.ph.ed.ac.uk}}

\abstract{We review simulations of dynamical domain wall fermions
at a fixed inverse lattice spacing of $1.73$GeV and with pion masses
as light as $330$MeV and spatial dimensions as large as $2.7$fm performed
by the RBC and UKQCD collaborations.
These results include pseudoscalar masses and decay constants
and low energy constants of the chiral effective lagrangian. 
We also review results for the neutral kaon mixing amplitude $B_K$,
the Kl3 form factor, pseudoscalar
meson structure, and vector meson decay constants. In the baryon sector we
review results for the spectrum, and nucleon form factors and structure functions.
Highlights of our programme include preliminary quark masses, and determinations
of $V_{us}$ from both $f_K/f_\pi$ and from Kl3, and an updated result
for $B_K$. We find significant finite volume effects in the nucleon axial
charge $g_A$ for our $m_\pi=330$ MeV ensemble on a $(2.7 {\rm fm})^3$ lattice,
and highlight the importance of large physical volumes for non-trivial nucleon
physics. 
}

\FullConference{The XXV International Symposium on Lattice Field Theory\\
		 July 30-4 August 2007\\
		 Regensburg, Germany}

\begin{document}

The RBC and UKQCD collaborations have jointly performed
2+1 flavour simulations of QCD, representing the
up, down, and strange quarks with the standard domain wall fermion action.
This review will firstly cover the theoretical foundations of our simulations
considering issues such as locality, chirality and topology. 
We will secondly present the ensemble parameters and algorithms used
in our simulations, and discuss algorithmic performance
and trade-offs.
We will finally summarise  important results presented elsewhere in this conference
\cite{nhc,ScholzLin,dja,cts,zanotti,ohta}
and recent publications
\cite{Allton:2007hx,Antonio:2007pb,Antonio:2007tr,Antonio:2006px,Boyle:2007wg,Antonio:2006ev,Antonio:2007mh,Boyle:2006xq,Boyle:2006pw}.
These results include pseudoscalar masses and decay constants
and low energy constants of the chiral effective lagrangian.
We also review results for the neutral kaon mixing amplitude $B_K$,
the Kl3 form factor, pseudoscalar
meson structure, and vector meson decay constants. In the baryon sector we
review results for the spectrum, and nucleon form factors and structure functions.
Highlights of our programme include preliminary quark masses, and determinations
of $V_{us}$ from both $f_K/f_\pi$ and from Kl3, and an updated result
for $B_K$.
We find significant finite volume effects in the nucleon axial
charge $g_A$ for our $m_\pi=330$ MeV ensemble on a $(2.7 {\rm fm})^3$ lattice,
and highlight the importance of large physical volumes for non-trivial nucleon
physics. 

\section{Lattice action, algorithms, cost}

We use the Iwasaki gauge action and domain wall fermion action
$$
  D^{\rm dwf}_{x,s; x^\prime, s^\prime}(M_5, m_f)
   = \delta_{s,s^\prime} D^\parallel_{x,x^\prime}(M_5)
   + \delta_{x,x^\prime} D^\bot_{s,s^\prime}(m_f)
$$
$$
D^\parallel_{x,x^\prime}(M_5)  = D_W(-M_5)
$$
\begin{eqnarray}
D^\bot_{s,s^\prime}(m_f) 
     &=& {1\over 2}\Big[(1-\gamma_5)\delta_{s+1,s^\prime} 
                 + (1+\gamma_5)\delta_{s-1,s^\prime} 
                 - 2\delta_{s,s^\prime}\Big] \nonumber\\
     &-& {m_f\over 2}\Big[(1-\gamma_5) \delta_{s, L_s-1}
       \delta_{0, s^\prime}
      +  (1+\gamma_5)\delta_{s,0}\delta_{L_s-1,s^\prime}\Big].
\end{eqnarray}
Here $D_W$ is the Wilson Dirac operator, and
the boundary conditions are understood to be Dirichlet in the
fifth dimension, periodic in spatial directions and anti-periodic in time.
Surface states of either chirality are bound to the 4-dimensional
$s=0$ and $s=Ls-1$ hyperplanes and are identifed
with physical, four-dimensional modes 
$$q(x) = P_L \Psi(x,0) + P_R \Psi(x,L_s-1).$$
In dynamical simulations the bulk infinity of the 
five dimensional partition function is removed using Pauli-Villars fields.

Our simulations, table~\ref{tabParameters},
have principally been performed using a single value for
the lattice spacing, $a^{-1}= 1.73$ GeV, and using both $16^3$ and $24^3$ lattice volumes
corresponding to $(2.0 {\rm fm})^3$ and $(2.7 {\rm fm})^3$.
A status report on  simulations in progress at a
second, finer lattice spacing with a $32^3$ lattice
is also given. These latter simulations are carried out 
as part of a collaboration between RBC, UKQCD, and LHPC.

\begin{table}
\begin{tabular}{|c|c|c|c|c|c|c|c|}
\hline
$L^3\times T\times L_s$ & $(a m_{l},m_s)$ & $\beta$ & $a^{-1}$ (GeV)& L (fm) &$m_\pi$ (MeV) &$m_{\rm res}$& $\tau$ MD \\
\hline
                        & $(0.01,0.04)$ &      &           &     & 400 &                     & 4000\\
$16^3\times 32\times 16$& $(0.02,0.04)$ &2.13  & 1.62(4)   & 1.94& 530 & $3.08\times 10^{-3}$& 4000\\
                        & $(0.03,0.04)$ &      & ($\rho$)  &     & 630 &                     & 7500\\
\hline
                        & $(0.005,0.04)$&      &           &     & 330 &                     &4500\\
$24^3\times 64\times 16$& $(0.01,0.04) $& 2.13 & 1.73(3)   & 2.73& 420 & $3.15\times 10^{-3}$&4700\\
                        & $(0.02,0.04)$ &      & ($\Omega^-$)&   & 560 &                     &2800\\
                        & $(0.03,0.04)$ &      &           &     & 670 &                     &2800\\
\hline
$32^3\times 64\times 16$& (0.004,0.03)&2.25&$\sim$2.15&$\sim$2.93&$\sim$260& $\sim 6\times 10^{-4}$&1100+\\
                        & (0.006,0.03)&    &          &          &$\sim$310&                     &1300+\\
\hline
\end{tabular}
\caption{Ensemble parameters for the UKQCD/RBC data set. The $32^3$ ensemble production is in
collaboration with LHPC since July 2007. The negative Wilson mass in the domain wall formalism was
1.8 for all ensembles.}
\label{tabParameters}
\end{table}

The degree of flexibility and choice of
the implementation of Hybrid Monte-Carlo has advanced greatly in recent
years, with several new algorithmic variants proposed. These advances include the
(affordable) extension of exact algorithms to odd numbers of flavours 
\cite{Clark:2004cp},
several schemes for splitting the fermionic force into UV and IR portions 
that can be updated on different timescales, and improved numerical 
integrators.
Experimenting  
with the available options\footnote{we 
gratefully acknowledge the 
immense contribution of Mike Clark to our programme}, 
we have settled on  RHMC with a hybrid combination multi-mass preconditioning 
at light mass scales 
\cite{Urbach:2005ji}
and multiple pseudofermion fields at
\cite{Clark:2006fx}
heavier mass scales with Omelyan integrators \cite{Takaishi:2005tz}.
The simulated fermion determinant is included as
$$
\det_A\left\{
\frac
{{D^\dag D}(m_{l})}
{{D^\dag D}(m_s)}
\right\}
\det_B\left\{
\frac
{{ D}(m_s)}
{{ D}(1)}
\right\}
\det_B\left\{
\frac
{{ D}(m_s)}
{{ D}(1)}
\right\}
\det_B\left\{
\frac
{{ D}(m_s)}
{{ D}(1)}
\right\},
$$
where each determinant factor is estimated via a separate pseudo-fermion field (thus four in all).
The degenerate $u,d$ flavours are mass preconditioned by the strange mass,
the remaining three factors of the strange mass make use of the RHMC n-roots
force reduction trick, and the factors of ${ D}(1)$ are the Pauli-Villars fields.
The ``A'' and ``B'' determinants are updated on different timescales using nested
Omelyan integrators with Omelyan parameter $\lambda = 0.22$. 

The coarsest timescale used is $\delta_\tau = \frac{1}{6}$ for the most expensive up/down fields
at our lightest mass. A trajectory length $\tau=1$ is used and thus contains only six timesteps.
The nature of the finer timesteps are somewhat complicated by our use of the Omelyan integrator.
An Omelyan integration QPQPQ timestep involves two 
force calculations that are not equally
spaced in Monte-Carlo time. Reversibility leaves little flexibility for
possible approaches to integrator nesting and in our
nomenclature a 1:1 nesting implements a \emph{complete} QPQPQ Omelyan timestep of the second
force contribution for each sub-timestep of the first. Our strange mass determinants
are 1:1 nested inside the up/down determinants implying that each of the three ``B'' force contributions
are calculated twice for every ``A'' force contribution. The gauge force is
nested inside the ``B'' contribution in a similar way but with an Omelyan nesting ratio of 1:6.
Convergence residuals used in molecular dynamics phases vary between $10^{-8}$ and $10^{-6}$ 
according to the typical force contribution, while $10^{-10}$ is used uniformly for all 
Metropolis steps. Guesses are history independent and 
reversibility has been demonstrated to very high precision.

The two most important measures of cost are the technology independent algorithmic cost and the
wall clock time to run an ensemble on the machines available to the collaboration. 
The scaling with light quark mass is weak for our mass-preconditioned algorithm.
To allow concrete comparisons with other calculations,
we quote that the $24^3$ ensemble with $m_{l}=0.02$ requires
$O(10^6)$ applications of $D_W$ for a $\tau=1.0$ trajectory. Around 40 $\tau=1$trajectories per day
are produced on a 4096 node QCDOC partition and sustained performance is around 1.1TFlop/s on this
machine size. For our $32^3$ simulations on the same machine size around 10
 units of MD time are produced per day, with a trajectory length of $\tau=2$ MD time units.

\section{Theoretical foundations}

The DWF five dimensional system can be represented as a Fock space trace with
a transfer matrix $T = e^{-H_T}$ where $\tanh \frac{H_T}{2} = \frac{H_W }{2+D_W} = K_S$. 
The four dimensional effective action of DWF is a functional 
of the gauge fields and is not manifestly local.
This approximates an overlap operator making use of $H_T$
as the argument to the sign function.
\begin{eqnarray}
\left[ \det D_{\rm dwf}(1)\right]^{-1} \det D_{\rm dwf}(m) &=&
 \det \frac{1}{2} 
\left[ 1+m + \gamma_5 (1-m) \tanh ( L_s \tanh^{-1} K_S) \right] \nonumber\\
&\to& \det \frac{1}{2} \left[1+m + (1-m)\gamma_5 {\rm sgn} K_S \right]
\end{eqnarray}

Ignoring anomalous chiral symmetry breaking for now, consider
the flavour non-singlet axial current in this formulation.
The five dimensional theory has a 
conserved five dimensional vector current. The DWF 
(five dimensional) axial transformation associates positive
and negative chiral charges with the positive and negative halves of the
fifth dimension. One can construct a four dimensional
axial current that is extensive in the fifth dimension and for which 
the chiral symmetry breaking effect of finite $L_s$ 
consists only of a mid-point term in the fifth dimension.
$$
 \Delta_\mu {\cal A}_\mu^a (x) = 2 m_f P^a(x) + J^a_{5q}(x)
$$
where 
$$
{\cal A}^b_\mu(x) = \sum_{s=0}^{L_s-1} {\rm sign}(s -\frac{L_s-1}{2})
j^b_\mu(x,s),
$$
$$
j^b_\mu(x,s) = \frac{1}{2}\left[
          \overline\Psi(x+\hat\mu,s)(1+\gamma_\mu)U^\dagger_{x+\mu,\mu} t^b \Psi(x,s)
         -\overline\Psi(x,s)(1-\gamma_\mu)U_{x,\mu} t^b \Psi(x+\mu,s)\right]
$$
$$
P^a(x) = \overline{\Psi}(x,0)P_R t^a\Psi(x,L_s-1)
       - \overline{\Psi}(x,L_s-1)P_L t^a\Psi(x,0) 
	          \equiv \overline{q}(x) \gamma_5 t^a q(x)
$$
$$
J^a_{5q}(x) = \overline{\Psi}(x,L_s/2)P_R t^a \psi(x,L_s/2-1) 
            - \overline{\Psi}(x,L_s/2-1)P_L t^a \psi(x,L_s/2) 
$$
In low energy Greens functions, the 
midpoint density $J^a_{5q}$ is equivalent to
the dimension-three operator $2 m_{\rm res} \overline{q} \gamma^5 q$ where
$m_{\rm res}$ is an additive mass renormalization, measured as
$$m_{\rm res} = \frac{\langle J^a_{5q}(x) P^a(y) \rangle}
                     {2 \langle P^a(x)        P^a(y) \rangle}.
$$
All unphysical chiral symmetry breaking effects in DWF, including
$m_{\rm res}$, involve propagation from a source
field of one or other chirality on the corresponding domain wall across
the fifth dimension. These naturally involve the
transfer matrix $T$ raised to an appropriate (large) power. 
This suppression mechanism is key to the quality
of DWF lattice chiral
symmetry, and thus understanding the
details of the nature of the spectrum of $H_T$ are crucial.
The asymptotic propagation in the fifth dimension is, for large $L_s$,
dominated by the modes of $H_T$ at the lowest eigenvalues at
which the eigenmode density $\rho(\lambda)$ is non-zero.
However, for modest $L_s$ these asymptotic contributions 
can be very much suppressed and details of $H_T$ and the
size distribution of the eigenmodes must be considered.

The translational invariance of the DWF approach in the 
fictitious fifth dimension admits power counting in $T^{L_s}$ as a powerful
tool, and $m_{\rm res}$ 
serves as qualitative guide to the cost of one factor of
$T^{L_s}$. It is worth emphasizing that this counting 
carries real power; for example wrong chirality mixings for $B_K$
require two crossings of the fifth dimension and
are proportional to $m_{\rm res}^2$ \cite{Christ:2005xh}.
Calculations of $\frac{\epsilon^\prime}{\epsilon}$ are feasible
\cite{Sharpe:2007yd,mawhinney}, while residual chiral
symmetry breaking is sufficiently enhanced in direct measurement
of the chiral condensate that it will prove problematic with DWF
without further reduction in $m_{res}$ (or matching
large $L_s$ valence  simulations to our sea pion masses)
\cite{Sharpe:2007yd}.

\subsection{Localisation structure of Aoki phase}

The zero modes of $H_T$ and $H_W$ coincide and the known quenched Aoki phase
behaviour of $H_W$ is directly relevant. It has recently been understood
that the structure of the quenched Aoki phase is rich: there is
a non-zero density of near zero modes of $H_W$, and a Banks Casher
pionic condensate, throughout all of the accessible phase diagram;
however, a localisation transition is thought to occur and, towards
weak coupling, the phase is thought to display a non-zero mobility edge.
Establishing that we at least have
a non-zero mobility edge (or better yet a gap in the spectrum)
is key to establishing locality of dynamical overlap or DWF simulations.
The conjectured model for the structure of the spectrum has $\rho(0)\ne 0$
and $\rho(\lambda_c) \gg \rho(0)$ - a volume factor enhances the contributions
from modes above the mobility edge $\lambda_c$.
A consequence of this model for the structure of the spectrum
of $H_T$ is that \cite{Antonio:2007tr,Golterman:2005fe}
$$
m_{\rm res}(L_s) = \frac{ c_1 +  c_2 e^{-\lambda L_s}}{L_s}.
\label{eq:mresls}
$$
Here these two contributions come from a low density of (volume factor
suppressed) localised near-zero modes and a larger density of 
extended modes near the mobility edge. The overall factor of
$\frac{1}{L_s}$ represents an infrared cut off on the shell of modes
that contribute significantly imposed by the size of the fifth dimension.

\subsection{Locality}
For sufficiently smooth gauge fields there is 
a gap in the spectrum of $H_W$ and this implies locality
of the corresponding overlap Dirac operator taking $H_W$ as the kernel of
the sign function \cite{Hernandez:1998et}.
This proof may be generalised to cover the Shamir
Kernel $K_S$ that corresponds to the $L_s\to\infty$ limit of DWF, and also
the condition can be relaxed to require only a gap in the spectrum of delocalised
eigenmodes \cite{Golterman:2003qe,Antonio:2007tr}.

As there is no gap in the spectrum of this $H_T$ (in
the absence of a ghost Wilson determinant) 
it is necessary to demonstrate that its spectrum displays a non-zero
mobility edge to establish locality of the effective four dimensional theory.
The related matrix $H_W = \gamma_5 D_W$ which has identical zero mode structure,
and it also suffices to study $H_W$ in its place. 

\begin{figure}
\begin{center}
\includegraphics*[width=0.6\textwidth]{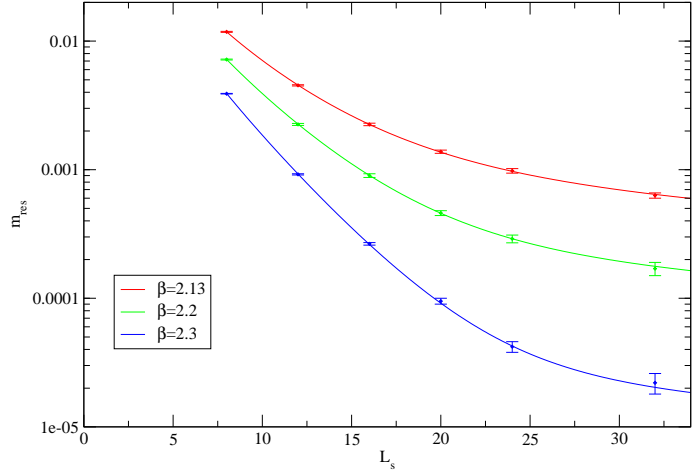}
\caption{We display the dependence of $m_{\rm res}$ on $L_s$ for valence
quarks on Domain Wall 2+1f ensembles with the Iwasaki gauge action and $L_s=8$.
The fit correspond to a model based on the conjectured mobility edge structure in
the Aoki phase, and a good description of our data is seen. This evidence
of a non-zero mobility edge indicates locality for the DWF effective action. }
\label{figMresLs}
\end{center}
\end{figure}

\begin{figure}
\begin{center}
\includegraphics*[width=0.6\textwidth]{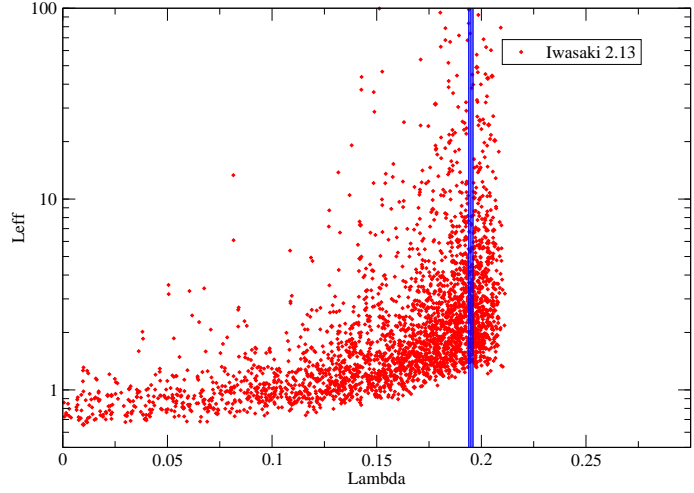}
\caption{We look at the scatter of exponential bounds measured for low modes
of the Hermitian Wilson Dirac operator for $M_5=-1.8$ for $\beta=2.13$. A mobility edge
is clearly seen, giving direct evidence of the locality of DWF for our simulated
parameters.}
\label{figLocalisation}
\end{center}
\end{figure}

This can be done in two ways. Firstly we have done so indirectly by demonstrating the consistency
of the behaviour of $m_{\rm res}(L_s)$ with the above model, figure~\ref{figMresLs}.
Secondly microscopic inspection of the eigenmodes of $H_W$ can check the locality of individual
low lying eigenmodes on a mode by mode basis, figure~\ref{figLocalisation}.
Given an eigenmode, $\psi(x)$, we take $y$ as the location of the 
maximum of $\psi^\dagger\psi(x)$ and find the lowest exponential localisation 
length $L_{\rm eff}$ that for \emph{all} coordinates $x$ with $|x-y| \ge 5$ satisfies bound
$$\psi^\dagger\psi (x) \le \psi^\dagger\psi (y) e^{-\frac{2 |x-y|}{L_{\rm eff}}}.$$
We note that the $L_{\rm eff}$ 
bounds the eigenmode large distance from its peak in \emph{all} directions,
and emphasize that this strict bound approach is robust against eigenmodes with
extended lower dimensional sub-spaces and other pathological cases. 
We therefore
have demonstrated a non-zero mobility edge $\lambda_c\ge 0.2$,
and thus locality, 
for our $\beta=2.13$ simulations and that we are therefore in a correct part
of the Aoki phase diagram for taking a continuum limit.

\subsection{Chiral symmetry breaking and non-perturbative renormalisation}

We use the Rome-Southampton RI-mom approach to determine the renormalisation
of our lattice operators non-perturbatively. The valence DWF action suppresses
$O(a)$ effects, both on and off shell, and thus is particularly well suited
to the off-shell renormalisation approach.
The good chiral symmetry of DWF is reflected in its renormalisation structure,
and one might expect this to be well demonstrated and tested by NPR.

\begin{figure}
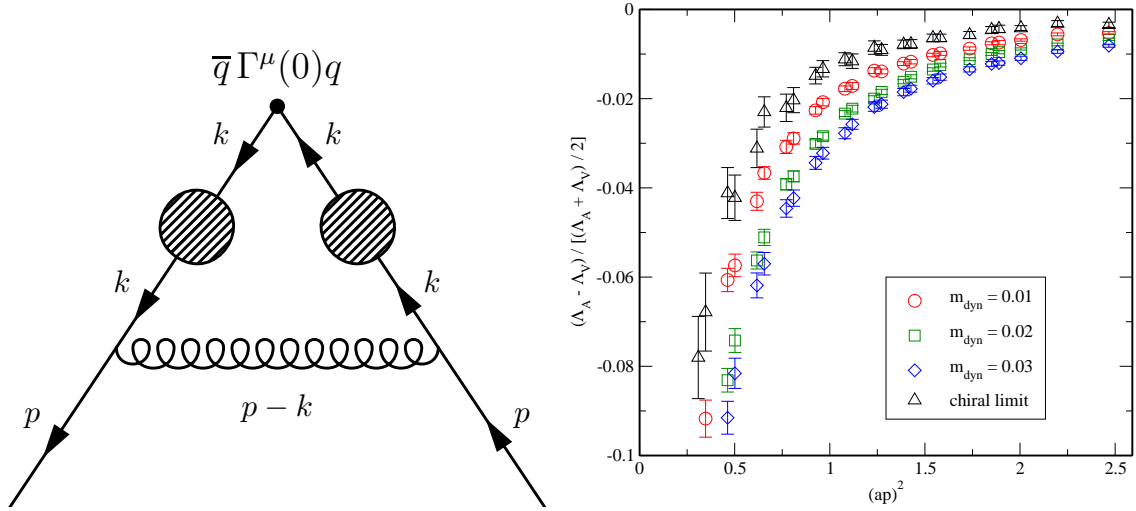

\includegraphics*[width=0.5\textwidth]{low-mom-chiral-breaking.epsi}
\includegraphics*[width=0.5\textwidth]{AVdiff.eps}
\caption{The left panel shows a class of vertex correction graphs admitting
non-perturbative physics accompanied by \emph{only} $\alpha_s/p^2$ suppression.
This is a disadvantageous feature of the standard RI-mom kinematics.
The right panel displays the (physical spontaneous) chiral symmetry breaking
effects that split the axial and vector amputated vertices $\Lambda_A-\Lambda_V$
for these kinematics.
The splitting is correspondingly poorly suppressed in $1/p^2$.}
\label{figNPRspontXSB}
\end{figure}

In practice, however, there are
substantial physical spontaneous chiral symmetry breaking effects at low $p^2$
for standard RI-mom kinematics.
These obscure a demonstration of the good chiral properties of DWF in coarse lattice
spacing simulations. An example of this for the symmetry breaking splitting of
$\Lambda_A-\Lambda_V$ in shown in figure~\ref{figNPRspontXSB}.
These effects also introduce an ambiguity in the determination 
of the renormalisation constants of around 2\%. This physical effect is
three orders of magnitude larger than any contamination 
expected from our residual chiral
symmetry breaking $m_{\rm res}^2 = (3\times 10^{-3})^2$ , and no improvement 
will be gained by improving the accuracy of the
chiral symmetry by either increasing $L_s$ or using a more exact overlap approach.

The problem has been enhanced by the particular choice of kinematics used in traditional RI-mom
NPR. For a standard bilinear vertex function the leg momenta
are equal and a soft subgraph is only suppressed by the $\frac{1}{p^2}$ of a single hard gluon
as shown in figure~\ref{figNPRspontXSB}.

\begin{figure}
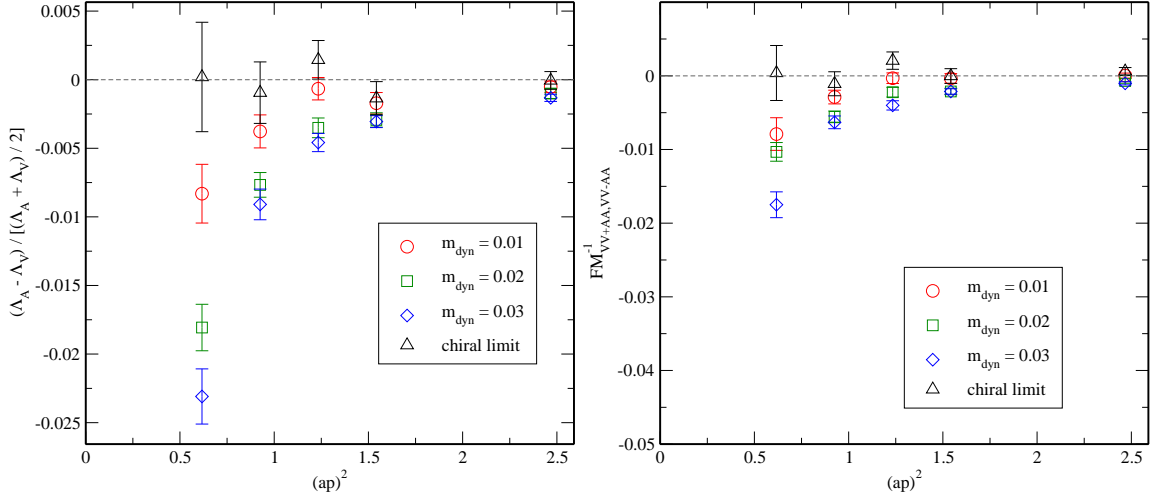

\includegraphics*[width=0.5\textwidth]{nonexp_AVdiff.eps}
\includegraphics*[width=0.5\textwidth]{Zbk_nonexp.eps}
\caption{By using non-exceptional momenta, spontaneous chiral
symmetry breaking effects are pushed to much lower momenta than
with standard RI-mom kinematics. 
The right panel displays
a cross-chirality mixing matrix elements relevant to $B_K$.
The excellent chiral properties
of DWF, when extrapolated to the chiral limit are now apparent
even at intermediate momentum scales. This approach is promising and
can demonstrate the absence of unwanted lattice mixings. In order to 
reduce NPR systematics for renormalised quantities
recalculation of RI-mom anomalous dimensions is required
to high order in continuum perturbation theory 
for these new kinematics. The left panel displays the difference between
the amputated axial and vector vertex functions. 
}
\label{fig:nonexceptional}
\end{figure}

A better alternative is to gain further suppression
of soft contributions using non-exceptional
momentum kinematics $p^2=p^{\prime2} = (p-p^\prime)^2$, figure~\ref{fig:nonexceptional}.

Unfortunately the large body of higher perturbative calculations for various
operators in RI-mom is non-trivial to reproduce for these alternative kinematics.
However, use can be made of these results since
even without conversion functions the good chiral chiral properties
of DWF can be demonstrated and omission of chirality mixing for $B_K$ justified
without requiring the perturbative conversion. 
There is no practical benefit, for these specific quantities, 
from further redution in $m_{\rm res}$. It is better to focus
our available effort on more pressing problems. 
The great improvement in the scaling window we have demonstrated for RI-mom vertex
functions with non-exceptional kinematics should serve as encouragement to 
the revisit high order calculations of the conversion
to $\overline{MS}$ for non-exceptional kinematics. It is also quite possible
that the convergence properties of the perturbative expansion will be improved
in some cases. For example the four-loop Wilson coefficient relevant to $Z_m$ displays much better
convergence in $\overline{MS}$ than for conventional RI-mom.

Table~\ref{tab:renorm} lists the axial current, field, mass, 
tensor and four quark operator
renormalisation constants obtained in reference~\cite{NPRinprep}.

\begin{table}\begin{center}
\begin{tabular}{|c|c|c|c|}
\hline
$\overline{\rm MS}$ 2 GeV      & Z & stat & sys\\
\hline
$Z_A$ & 0.7161& 0.0001 & \\
$Z_m$ & 1.656 & 0.048  & 0.104 \\
$Z_T$ & 0.7951& 0.0034 & 0.0117\\
$\frac{Z_{{\cal O}_{VV+AA}}}{Z_A^2}$ & 0.9276& 0.0052 & 0.0222\\
\hline
\end{tabular}
\caption{
Renormalisation constants for our $\beta=2.13$ ensembles. Results are 
quoted in the
chiral limit, and obtained on the $16^3$ ensembles. Systematic errors quoted contain
estimates of extrapolation errors and those from (continuum) perturbative conversion
to $\overline{MS}$. The exception is $Z_A$ which is an improved
measurement of the ratio of the conserved axial current ${\cal A}_\mu(x)$
to the more commonly measured boundary field axial current 
$\overline{q}(x)\gamma_5\gamma_\mu q(x)$.
}\label{tab:renorm}
\end{center}
\end{table}

\subsection{Topological tunneling}

\begin{figure}
\begin{center}
\includegraphics*[width=0.6\textwidth]{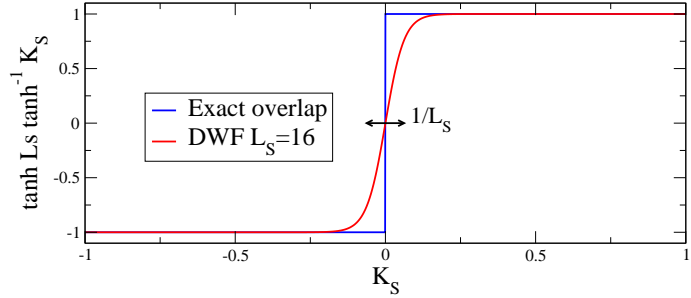}
\caption{Tanh approximation to the sign function for $L_s=16$}
\label{figStep}
\end{center}
\end{figure}

\begin{figure}
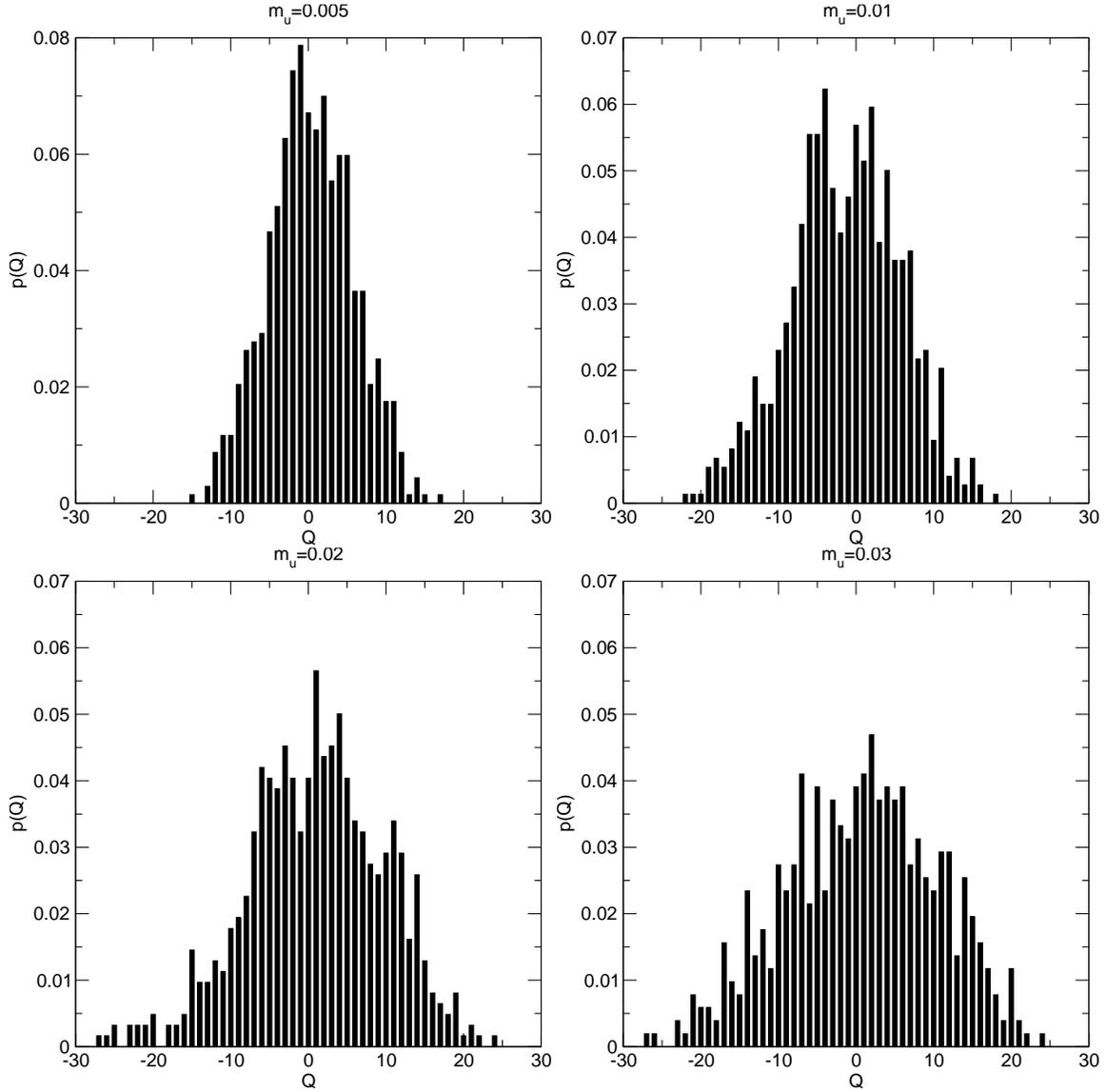

\includegraphics*[width=0.5\textwidth,angle=-90]{0.005.epsi}
\includegraphics*[width=0.5\textwidth,angle=-90]{0.01.epsi}\\
\includegraphics*[width=0.5\textwidth,angle=-90]{0.02.epsi}
\includegraphics*[width=0.5\textwidth,angle=-90]{0.03.epsi}
\caption{Topological charge distribution on our four $24^3$ ensembles.
These are $m_{l} = 0.005$ (top-left), $0.01$ (top-right), $0.02$ (bottom-left),
and 0.03 (bottom-right). The light fermion mass is clearly constraining the distribution,
and we are likely sampling topology well enough to reproduce $\theta=0$ QCD.}
\label{figTopoHist}
\end{figure}

\begin{figure}
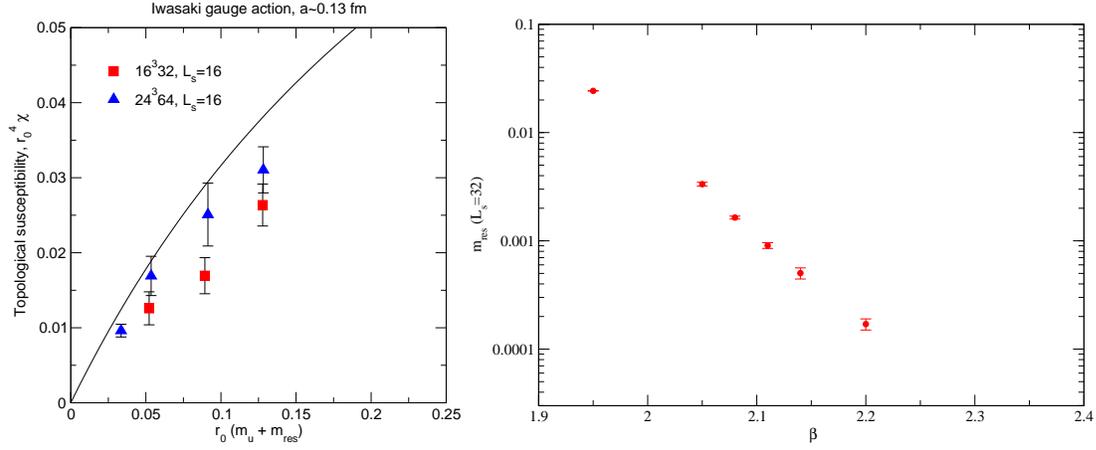

\includegraphics*[width=0.4\textwidth]{khi_mu_07.eps}
\includegraphics*[width=0.55\textwidth]{mresl32.eps}
\caption{Left panel displays topological susceptibility on our $24^3$ and $16^3$
configurations; the line is the leading order chiral behaviour with $\Sigma$ taken from
the Gell-Mann-Oakes-Renner relation.
Right panel displays the dependence of $m_{\rm res}(L_s=32)$ as a function of inverse gauge
coupling. As discussed in the text this can be taken as loosely indicative of trends in the 
density of near zero modes and also of the topological tunneling rate per unit lattice volume.}
\label{figTopSusc}
\label{figMresl32}
\end{figure}

One of the principal attractions of dynamical fermion simulations with good
chiral symmetry is the existence of a correct axial anomaly. 
However, this perfection of the action could easily be compromised 
by mundane algorithmic issues resulting in failure to sample the topological
structure of the vacuum adequately. Problems arise with exact chiral 
symmetry since the sign function has a discontinuity which molecular
dynamics updates will skip over for any non-zero timestep. 
Approximations to the sign function, DWF included, can involve a smooth
transition over some eigevalue range, figure~\ref{figStep}. 
Problems of this nature have not been observed with DWF simulations, and are unlikely to appear
as the integration is problem free provided
$
\delta\tau \ll \frac{1}{L_s\dot{\lambda}}.
$
Healthy global topological charge histograms are obtained on our $24^3$ ensembles 
in figure~\ref{figTopoHist} and the susceptibility is displayed in figure~\ref{figTopSusc}.

The molecular dynamics problems have resulted in two responses from the dynamical
overlap community. The reflection-algorithm treats the lowest modes exactly using
a particularly expensive approach, while others \cite{Matsufuru} have used an auxiliary pseudofermion determinant
to freeze the global topological charge by suppressing the density of low modes of $H_W$.
Here, it is worth noting that the Zolotarev approximation to the sign function used
has a coarse lower bound $|\lambda|\ge 0.1$ and does not differ 
greatly from the $\tanh$ displayed in figure~\ref{figStep}.
In the language of the overlap, the key difference from the RBC-UKQCD approaches is 
that the authors of \cite{Matsufuru} suppress the density of low modes of 
$H_W$ in the region $-0.1 \le \lambda \le 0.1$ using an auxiliary fermion
determinant, and it is then feasible to project and treat exactly the few 
remaining low eigenmodes of $H_W$. This results
in a fixed topology simulation with improved chiral symmetry.
Similar approaches could equally well be combined with the DWF $\tanh$ 
approximation \cite{Vranas:1999rz,Vranas:2006zk,Izubuchi:2002pq,Edwards:2000qv}.

We estimate that DWF simulations are around five times cheaper than the five dimensional
Zolotarev approach used in \cite{Matsufuru}, and twenty times cheaper than the
nested four dimensional approach.

As DWF shows, issues with integrability of the fermion contributions in molecular dynamics
are likely either solvable or avoidable. We note that were the auxiliary determinant and projection dropped, 
then the remaining Zolotarev approximation would yield 
a very ``domain-wall-like'' simulation
and tunneling would likely take place. 
Some thought has recently been given to developing
algorithms that continue to tunnel topology in the presence of an auxiliary determinant 
\cite{Golterman:2007ni}.

A more fundamental problem is the increasing potential barrier introduced between topological
sectors by the gauge action with increasing $\beta$. Eq.~\ref{eq:mresls} and discussion
indicates that $m_{\rm res}(L_s=32)$ can be a qualitative guide to the near-zero mode density $\rho(0)$.
Figure~\ref{figMresl32} shows that this, and we conjecture the tunneling rate per unit lattice volume,
vanishes exponentially in the gauge coupling.

The trade off is clear: suppressing the low modes that mediate topology change
and attempting to answer the difficult question about ergodicity of the simulation is one rational
choice. RBC and UKQCD's choice has been to
accept a level of residual chiral symmetry breaking that is merely a minor irritant, but avoid
questionable ergodicity and
the risk of getting anomaly physics wrong in a particularly expensive fashion. This is also  
a substantially cheaper simulation, allowing more important systematic issues to be 
addressed.

\section{Measurements}

\begin{table}
\begin{tabular}{|c|c|c|c|c|c|c|}
\hline
 mx,my & 0.001 & 0.005 & 0.01 & 0.02 & 0.03 & 0.04 \\
\hline
0.001 & 
  {\color{blue}A},B ;
  {\color{red}240}  
      &
  {\color{red}290}   
      &
  {\color{red}340}   
      & 
  {420}   
      &
  {490}   
      &
  {550}   \\
0.005 &       
  {A,B}
      & 
  {{\color{blue}A},B,{\color{green}C}};
  {\color{red} 330} 
       &
  {\color{red} 370} 
       &
  {450} 
       &
  {520} 
       & 
  {580}   \\
0.01  &       
      A,B
      &       
      A,B
      & 
  {{\color{blue}A},B,{\color{green}C}};
  {\color{red} 410} 
       &
  {480} 
       &
  560 
       &
  600   \\
 0.02  & A,B    & A,B      &  A,B    & 
  {\color{blue} A},B,{\color{green}C};
  {550}   
       &
  {600}   
       &
  {650}   
      \\
 0.03  &  A,B   &  A,B   & A,B   & A,B     & 
  {\color{blue}A},B,{\color{green}C};
  {650}   
       & 
  700  
       \\
 0.04  &  A,B    & A,B,C     & A,B,C     & {\color{blue}A},B,{\color{green}C}     &  {\color{blue}A},B,{\color{green}C}   & 
  {\color{blue}A},B,{\color{green}C};
  750   \\
\hline
\end{tabular}
\caption{
We display the available mesonic measurements for the four $24^3$ ensembles $m_{l}\in \{ 0.005,0.01,0.02,0.03\} ; m_s=0.04$. The table
lists valence masses $m_x$ and $m_y$. In the upper right triangle we give the approximate pseudoscalar meson mass in MeV composed of quarks with valence
masses $m_x$ and $m_y$. Those masses quoted in red survive our cuts for making NLO partially quenched chiral fits.
In the lower left triangle we denote meson mass and decay constant measurements by ``A'', neutral meson mixing matrix element measurements by ``B'', and
semileptonic decay matrix elements and distribution amplitudes by ``C''. 
Black corresponds to valence measurements made only on the lightest two ensembles.
Blue corresponds to valence measurements made on all four ensembles.
Green corresponds to unitary measurements made only with the valence quark masses equal to sea quark masses.
}
\label{tabMeasurements}
\end{table}

Table~\ref{tabMeasurements} gives a summary table of the mesonic measurements made on our
ensembles. 
Further measurements have been made of nucleon two and three point functions, 
and static-light two and three point functions \cite{jan,yasumichi}. 
Between 150 and 700 measurements have been made on each ensemble
depending on the quantity, and valence pseudoscalar masses vary between 240 MeV and 750 MeV. 
Dynamical pion masses run from 330MeV to 650 MeV. Many more valence masses than dynamical
masses are used to increase the amount of information in the chiral regime
and are exploited in fits to partially quenched chiral perturbation theory \cite{Sharpe:2001fh}, as highlighted by
the masses quoted in red.

\section{Chiral effective lagrangian}

In this section we review the results presented in more detail 
by Meifeng Lin and Enno Scholz at this conference \cite{ScholzLin}.
Two approaches to fitting our data for pion and kaon masses and 
decay constants to obtain the 
LEC's of the chiral effective Lagrangian have been presented
at this conference. 
The finite range of validity of chiral perturbation theory 
leaves such determinations
from lattice (or indeed real world) data as something of an art. 
The rather massive real world kaon is neither unambiguously 
light nor heavy compared with chiral scales.

One approach is to fit the full SU(3)$\times$SU(3) chiral effective theory to data including the kaon
as an active chiral pseudoscalar. 
A strength of lattice formulations, such as DWF, with controlled flavour symmetry
is that the chiral perturbation theory can be decoupled from 
lattice artefacts in our simulated flavour content.

A good available alternative is to treat the kaon as a 
non-Goldstone boson, coupled to an effective SU(2)$\times$SU(2) theory. 
The analysis is applicable whenever $m_\pi \ll m_K$, 
whether or not the kaon is heavy or light compared
to other scales and we do not rely on chiral perturbation 
theory being convergent at kaon masses. 
The LEC's will be strange mass dependent and,
since the kaon is somewhat lighter than a typical chiral scale, the
convergence of the chiral expansion controlled by these LEC's 
may be correspondingly impacted.
This will merely reflect the new dynamics that enters at the kaon mass scale.
A reasonable estimate is that successive orders in the chiral 
expansion will \emph{only} be suppressed by $m_{l}/m_s$.
To estimate possible systematic NNLO contamination entering 
when we perform SU(2) fits with a mass cut-off $m_l$ 
we multiply the size of NLO corrections in our 
NLO fit by $m_l/m_s$.
The relevant formulae can be easily obtained from standard SU(2) heavy meson 
chiral perturbation theory 
under the simplification that vector contributions are 
dropped\cite{Sharpe:1995qp}. 

We find that SU(3)$\times$SU(3) NLO partially quenched chiral perturbation theory 
\cite{Sharpe:2001fh}
does not describe our data well up to meson masses comparable to the kaon mass. A good fit
can only obtained with a cut in bare quark mass of $a m_{\rm avg} \equiv (a m_x + a m_y)/2 \le 0.01$.
while the kaon corresponds to $a m_{\rm avg} \simeq 0.016$.  The utility of SU(3)$\times$SU(3) ChPT at NLO is questionable; however we can quote LEC's that 
fit the data for valence strange and light quark masses 
obeying the bound $a m_{\rm avg}\le 0.01$,
and light dynamical quark masses $a m_{l}\le 0.01$
but with a fixed dynamical strange quark mass of $a m_s = 0.04$.
These LEC's may well differ from those that would be 
obtained in the unphysical true SU(3) chiral limit.
The low energy constants obtained are displayed in table~\ref{tab:fitres} 
for two popular choices of the chiral scale.

\begin{table}
%
%
\caption{\label{tab:fitres}Fitted parameters from different fits with a valence mass cut $am_{\rm avg}\leq0.01$. For each fit the LECs are quoted at two diffe
rent scales $\Lambda_\chi$. (Note: the value of $B_0$ depends on the renormalization scheme like the quark masses: to obtain $B_0$, e.g., in the $\overline{\rm MS}(2\,{\rm GeV})$ scheme, one has to divide the values quoted here
by $Z_m^{\overline{\rm MS}}(2\,{\rm GeV})$.} 
\begin{center}
\begin{tabular}{cR*{4}{C}}\hline
&\multicolumn{1}{C}{\Lambda_\chi} & (2L_8-L_5) & L_5 & (2L_6-L_4) & L_4 \\[3pt]\hline
\multicolumn{6}{L}{\SU(3)\times\SU(3):\:aB_0 = 2.35(16),\, af_0 =  0.0541(40)}\\[2pt] 
                   & 1\,{\rm GeV}
                   & 5.19(45)\sci{-4} & 2.51(99)\sci{-4} & -4.7(4.2)\sci{-5} & -6.7(8.0)\sci{-5} \\
                   & 770\,{\rm MeV}  
                   & 2.43(45)\sci{-4} & 8.72(99)\sci{-4} & -0.1(4.2)\sci{-5} & 1.39(80)\sci{-4} \\[5pt]
\multicolumn{6}{L}{\SU(2)\times\SU(2):\: aB_0 = 2.414(61),\, af_0 =  0.0665(21)}\\[2pt] 
                   & 1\,{\rm GeV}  
                   & 4.64(43)\sci{-4} & 5.16(73)\sci{-4} & -7.1(6.2)\sci{-5}&  1.3(1.3)\sci{-4}\\
                   & 770\,{\rm MeV}  
                   & 5.0(4.3)\sci{-5} & 9.30(73)\sci{-4} & 3.2(6.2)\sci{-5} & 3.3(1.3)\sci{-4} \\\hline
\end{tabular}
\end{center}
\end{table}

We also directly applied SU(2)$\times$SU(2) to fit the LEC's of the effective two flavour theory that matches
our simulated 2+1 flavour ``real'' world. These are also displayed in table~\ref{tab:fitres}.
We fit the NLO forms to our data using a mass cut $am_{\rm avg}\leq0.01$ to obtain a good
quality of fit.
We perturbatively convert our results to the scale independent $\bar{l}_3$, $\bar{l}_4$
SU(2) LEC's in table~\ref{tab:conv_SU3_SU2}. 
Interestingly our 2+1f results, both from a perturbatively converted
SU(3) fit and from a direct SU(2) fit are broadly consistent
with each other and with the 2f results of ETMC
\cite{Boucaud:2007uk}
and CERN
\cite{DelDebbio:2006cn,DelDebbio:2007pz}. 
This adds somewhat to the picture discussed 
recently by Leutwyler \cite{Leutwyler:2007ae}.

\begin{table}
\caption{\label{tab:conv_SU3_SU2}Comparison of converted $\SU(3)\times\SU(3)$ fit parameters with those from $\SU(2)\times\SU(2)$ fits. Low energy scales $\bar{l}_{3,4}$ are defined at $\Lambda=139\,{\rm MeV}$.}
\begin{center}
\begin{tabular}{l*{3}{c}c}\hline
                               &$aB_0$ & $af_0$ & $\bar{l}_3$ & $\bar{l}_4$ \\\hline 
$\SU(3)\times\SU(3)$, conv. & 2.457(78) & 0.0661(18) & 2.87(28) & 4.10(05) \\
$\SU(2)\times\SU(2)$ & 2.414(61) & 0.0665(21) & 3.13(33) & 4.43(14) \\\hline 
\end{tabular} 
\end{center}
\end{table}

\section{Quark masses and lattice spacing}

We determine
$a^{-1}, a m_{ud}$  and $a m_s$
from a combination of the S=3 $\Omega^-$ baryon mass and
the pseudoscalar kaon and pion masses. These quantities then
produce the correct $\Omega^{-}$ mass from linear extrapolation
in the valence mass to $a m_s$ and to $a m_{ud}$ in the light sea masses. They
simultaneously produce consistent kaon and pion masses 
using our SU(2)$\times$SU(2) chiral extrapolations.
\begin{table}
\caption{\label{tab:results}Determined lattice scale and spacing and unrenormalized quark masses ($am_x^{\rm phys}=am_x^{\rm bare}+am_{\rm res}$).}
\begin{center}
\begin{tabular}{*{6}{C}}\hline
  a^{-1}/{\rm GeV} & a/{\rm fm} & am_{ud}^{\rm bare} & am_{ud}^{\rm phys} & am_s^{\rm bare} & am_s^{\rm phys} \\\hline 
  1.729(28) & 0.1141(18)  & -0.001847(58) & 0.001300(58)  & 0.0343(16)   & 0.0375(16)\\\hline 
\end{tabular}
\end{center}
\end{table}
The scale from the $\Omega^-$ can also be used to predict $f_\pi$, $f_K$, 
$\frac{f_K}{f_\pi}$ and the quark masses  \cite{ScholzLin}.
Based on our preliminary analysis 
our bare quark masses correspond to renormalised quark masses 
of  $m_{ud}=3.72(16)\,{\rm MeV}$ and  $m_s=107.3(4.5)\,{\rm MeV}$
in the $\overline{\rm MS}$ at 2 GeV. We find $\frac{m_s}{m_{ud}} = 28.8(4)$.
We obtained $f_\pi = 124.1(3.6)$ MeV, $f_K=149.6(3.6) {\rm MeV}$, which are around 5\% lower than their
experimental values. This is likely an $O(a^2)$ effect, and our ratio $\frac{f_K}{f_\pi} = 1.205(18)$.
This implies $|V_{us}| = 0.2232(34) $. 
Here the errors on decay constants are statistical only,
and the quark masses and $V_{us}$ contain only partial systematic errors.
Full systematic errors will be estimated in a journal paper
\cite{24inprep}.

The lattice spacing determined from $M_\rho$ is somewhat different,
being around 1.65 GeV on $24^3$ and 1.62GeV on $16^3$. 
While on $16^3$ we relied on  vector meson states (which
are unstable in QCD) and an ad hoc value of $0.495$ fm for $r_0$ to set the 
scale, our larger $24^3$ volume enables $f_\pi$ and 
baryon masses to be safely considered.
We find broad consistency between decay constants 
and the Omega (which, being composed of three
strange quarks, is physically small for a baryon). 
We have found that in the chiral limit $r_0/a= 4.13(10)$, and this suggests
we measure a physical value of $r_0 \simeq 0.47$ fm, and disfavour $r_0 = 0.495$ fm.
Were we to use the pseudoscalar decay constants to set the scale $r_0 \simeq 0.45$ fm.
This is a tendency that is consistent with other recent lattice calculations \cite{mcneile},
and cautions against reliance on vector mesons for precision scale. We are encouraged 
by recent progress that has been made on treating vector meson decay
in lattice QCD \cite{Aoki:2007rd}.

\section{Neutral kaon mixing}

We have updated our paper
\cite{Antonio:2007pb} on $B_K$ with $24^3$ results that have been 
presented by Cohen and Antonio at this conference \cite{dja}. 
We use the two wall, operator sink method to gain a spatial volume average.
We use propagators that are the sum of solutions for periodic and anti-periodic
temporal boundary conditions to eliminate unwanted round-the-world propagation.
This gives exceedingly long plateaux on our $N_t=64$ lattice.
We have modified our analysis to set the lattice spacing 
from the $\Omega^-$ mass and now use fits
assuming only SU(2)$\times$SU(2) chiral symmetry which we consider to
be more theoretically robust. We have access to lighter
masses and more statistically precise data 
and see evidence of curvature in the fixed strange mass chiral 
extrapolation. Partially quenched SU(2)$\times$SU(2) 
chiral perturbation theory both describes the valence and sea 
mass dependence well in the region of our fit.
The smaller volume $m_{l}=0.02$ data point 
is not included in our fit, but the unitary
fit curve matches onto this data point reassuringly well.

\begin{figure}
\begin{center}
\includegraphics[width=0.6\textwidth]{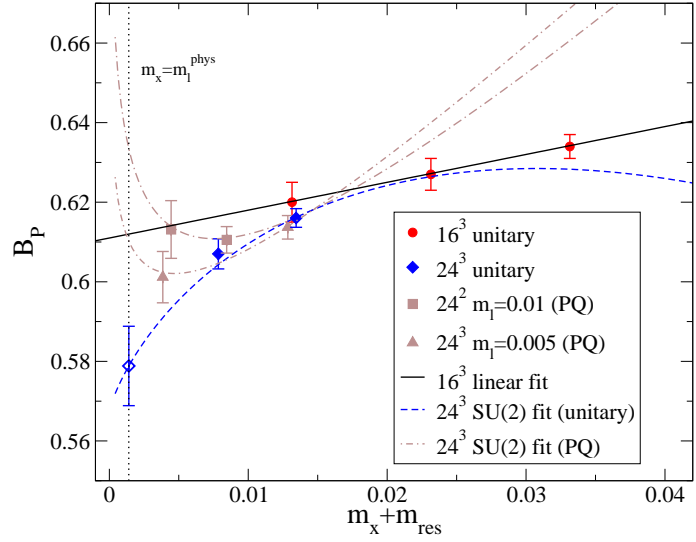}
\end{center}
\caption{\label{fig:chiral}
Results for $B_P$ together with the NLO partially quenched 
$SU(2)\times SU(2)$ ChPT fit to the $24^3$ data plotted versus the light valence quark 
mass $m_x$.  From top to bottom on the left-hand-side, the three curves 
are $m_l$ = 0.01, 0.005 and $m_x$ respectively.  The valence strange 
quark mass is fixed at its unitary value $m_y = m_s = 0.04$.  While 
the statistical errors are large, the growing upward curvature in 
$m_x$ as the sea quark mass is increased from 0.005 to 0.01 
predicted by ChPT is visible.  The $m_x$ values are slightly 
shifted for clarity.}
\end{figure}

The final results follow, with the first error statistical and the
second systematic.
\begin{eqnarray}
B_K^{\rm RI}(2 \mbox{ GeV})             &=& 0.514(10)(25), \\
B_K^{\overline{\rm MS}}(2 \mbox{ GeV})  &=& 0.524(10)(28),  \\
\hat{B}_K                               &=& 0.720(13)(37),
\end{eqnarray}

The components of the systematic error are shown in table~\ref{tab:bksyserr}.
These errors are added in quadrature, and the discretisation systematic is dominant.
Simulations in progress with a finer lattice 
spacing will directly address this. 
The two-loop perturbative conversion to $\overline{MS}$ is currently a 
subleading error but will soon become the most 
important error to address. Finer
lattice spacings will only yield logarithmic improvement, 
and a higher order calculation, preferably with non-exceptional momenta, 
is important. A non-perturbative step scaling approach could, of course,
even better address the convergence of perturbation theory.

\begin{table}
\begin{center}
\begin{tabular}{cc}
Non-perturbative renormalisation  & 2\% \\
Sea strange mass adjust & 1\% \\
Chiral extrapolation & 2\% \\
Discretisation       & 4\% \\
Finite volume        & 1\% 
\end{tabular}
\caption{Breakdown of systematic error estimate for our $24^3$ $B_K$ result.}
\label{tab:bksyserr}
\end{center}
\end{table}

\section{Kl3 form factor}

\begin{figure}
\includegraphics*[width=0.5\textwidth]{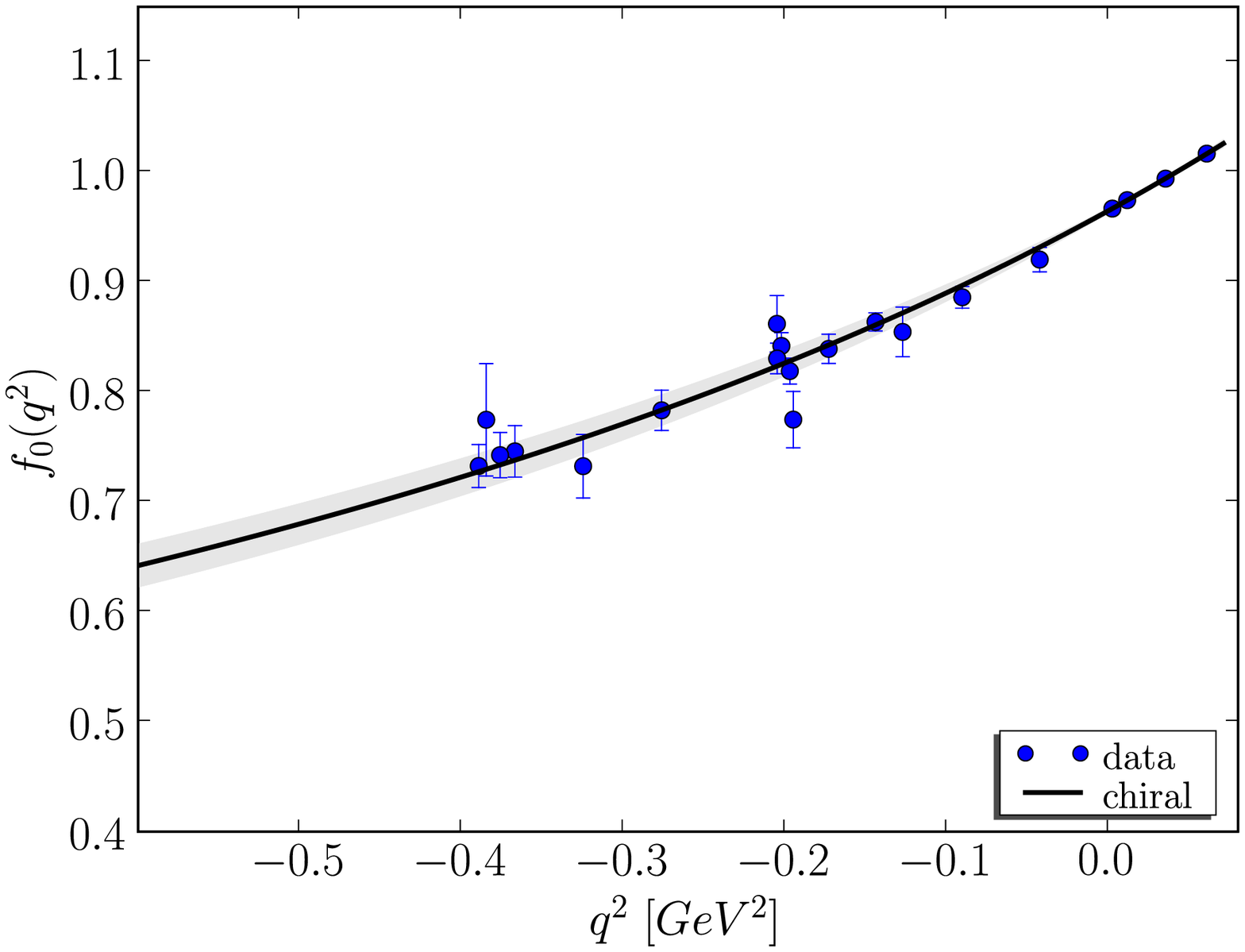}
\includegraphics*[width=0.5\textwidth]{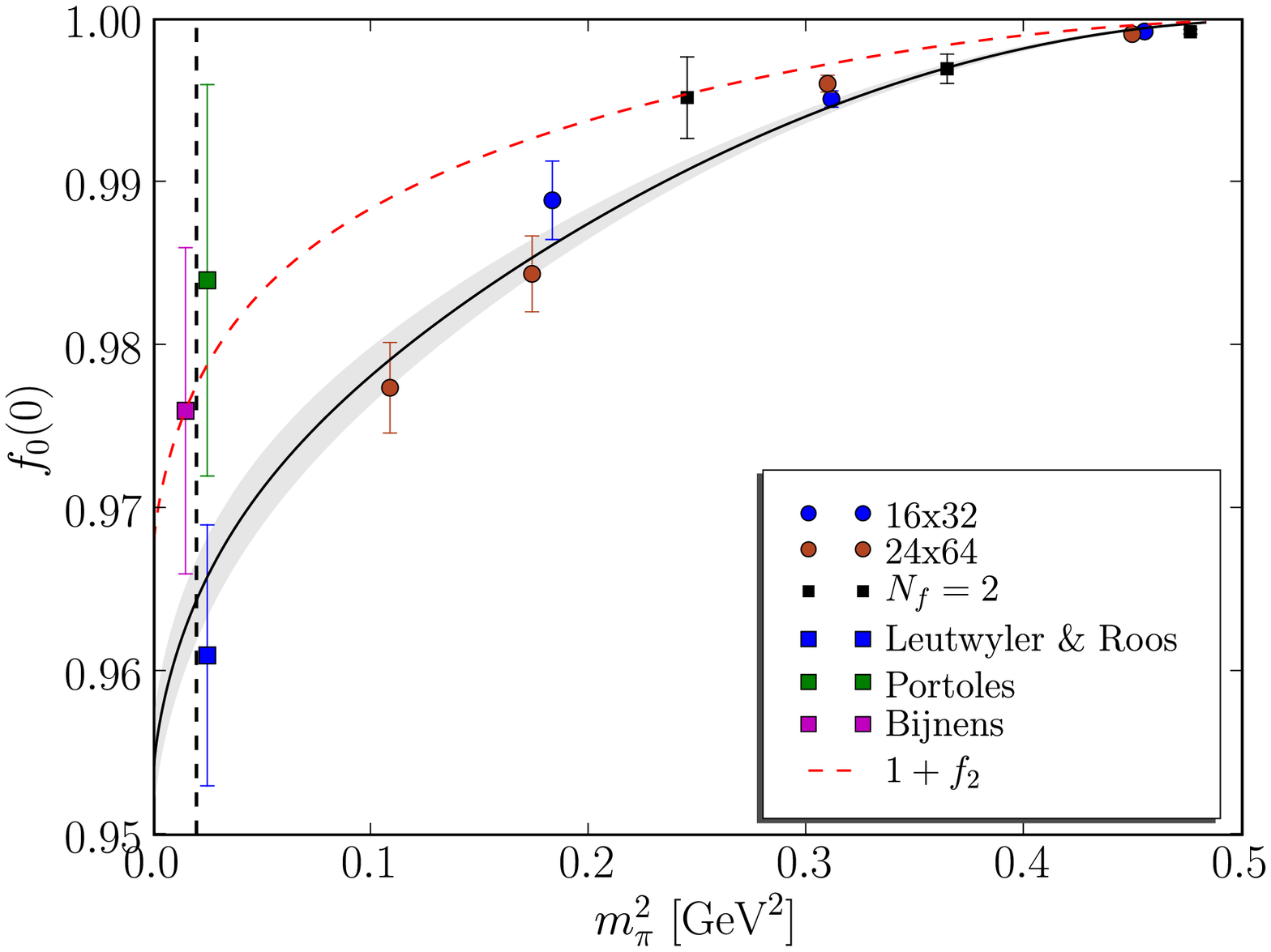}
\caption{ We use a unified fit on the $24^3\times 64$ data to both the
  $q^2$ and mass dependence of the form factor.  The data can be
  usefully displayed in two ways.  Left panel contains a pole
  dominance model interpolation of the form factor $f_0(q^2)$ to $q^2
  = 0$ having extrapolated to the chiral limit; the data points are
  adjusted, using the fit model, such that if the fit were perfect
  they would all lie on the fit model. The small remaining scatter is
  an indication of the quality of our unified fit.  The right panel
  shows the chiral extrapolation of $f_0(q^2=0)$ ; here the line is
  our fit model, while the data points are the results of
  interpolation to $q^2=0$ for each ensemble and these are
  consistent. We favour the Leutwyler-Roos prediction and have a
  smaller error.  }
\label{kl3fig}
\end{figure}

James Zanotti presented a status update of our calculation of the 
semileptonic kaon decay form-factor\cite{zanotti}, 
$f_+(0)$ which is obtained from
the $K\to \pi$ matrix element of the weak vector current
$$ \langle \pi(p^\prime)|V_\mu|K(p)\rangle 
= 
f_+(q^2) (p_\mu+p^\prime_\mu) + f_-(q^2) (p_\mu-p^\prime_\mu) 
$$
This is a promising approach for an accurate determination of $V_{\rm
  us}$, and makes use of standard double ratio techniques
\cite{Becirevic:2004ya} to measure the deviation of the form factor
from unity, giving a very small overall error. An example ratio is
given below.
$$ 
\frac{\langle K(\vec{0}) | V_0 | \pi(\vec{0}) \rangle \langle
  K(\vec{0}) | V_0 | \pi(\vec{0}) \rangle} 
     {\langle K(\vec{0}) | V_0 | K(\vec{0}) \rangle \langle
       \pi(\vec{0}) | V_0 | \pi(\vec{0}) \rangle} 
= \frac{(m_K+m_\pi)^2}{4m_K m_\pi} |f_0(q^2_{\rm max})|^2
$$
Preliminary results using only the $m_u=0.03,\,0.02,\,0.01$ data
points were presented in \cite{Antonio:2007mh,Antonio:2006ev}.
We have now added the lightest data point ($m_u=0.005$) and finalised
our analysis in a full paper \cite{Boyle:2007qe}.
The updated analysis includes unified chiral and $q^2$ extrapolations,
using a fit form that combines a quark mass dependent pole dominance
model with the constraints of the Ademollo-Gatto theorem:
\begin{equation}
f_0(q^2,m_\pi^2,m_K^2) = 
\frac{1+f_2+(m_K^2-m_\pi^2)^2 ( A_1+A_2(m_K^2+m_\pi^2) )}
{1 - \frac{q^2}{(M_0 + M_1 (m_K^2+m_\pi^2))^2} }\, .
\label{eq:globalpole}
\end{equation}
The results from a fit to the large volume ($24^3$) data is presented
in Fig.~\ref{kl3fig}.
The left and right plots show the $q^2$ and quark mass dependencies of
Eq.~(\ref{eq:globalpole}), respectively.
At the physical meson masses, we obtain 
\begin{equation}
 f_+^{K\pi}(0) = 0.9644(33)(34)(14)\ ,
\label{eq:fzero}
\end{equation}
which very much favours Leutwyler-Roos \cite{Leutwyler:1984je} results
over more recent higher order calculations
\cite{Bijnens:2003uy,Cirigliano:2005xn}.
The first error in Eq.~(\ref{eq:fzero}) is statistical, while the
second is an estimate of the systematic error due to our choice of
ansatz (\ref{eq:globalpole}) and the third is the estimate of
discretisation errors.
The PDG quotes $|V_{us}f_+(0)|=0.2169(9)$ \cite{blucher06}\footnote{A more recent analysis finds $|V_{us}f_+(0)|=0.21673(46)$ \cite{Moulson:2007fs}}, 
so using our result (\ref{eq:fzero}),
we obtain
$$ 
|V_{us}| = 0.2247(9)_{\rm exp}(11)_{f_+(0)}\ .
$$
Despite being less mature, Kl3 form factor appears very competitive
with respect to $f_K/f_\pi$ as a lattice method for constraining
$V_{us}$.
We anticipate a substantial reduction in error by a recalculation with
a combination of twisted boundary conditions \cite{Boyle:2007wg}
(removing the systematic uncertainty in the $q^2$ extrapolation) and
stochastic volume averaging for our $m_u=0.005$ data point (reducing
the error in the most important point in the chiral extrapolation).
Finally, discretisation effects will be addressed before the next
lattice conference using our new ensembles with a finer lattice
spacing.

\section{Pion and kaon distribution amplitudes}

\begin{figure}
\begin{minipage}{.48\linewidth}\begin{center}
\psfrag{mqplusmres}[t][c][1][0]{\footnotesize$am_q+am_{\rm res}$}
  \psfrag{scndmom}[b][c][1][0]{\footnotesize$\xisqav_\pi^{\rm bare}$}
  \psfrag{Legendmass1}[c][c][0.8][0]{\footnotesize$am_l=0.01$}
  \psfrag{Legendmass2}[c][c][0.8][0]{\footnotesize$am_l=0.02$}
  \psfrag{Legendmass3}[c][c][0.8][0]{\footnotesize$am_l=0.03$}
  \psfrag{Legendmass4}[c][c][0.8][0]{\footnotesize$am_l=0.005$}
  \epsfig{scale=.25,angle=270,file=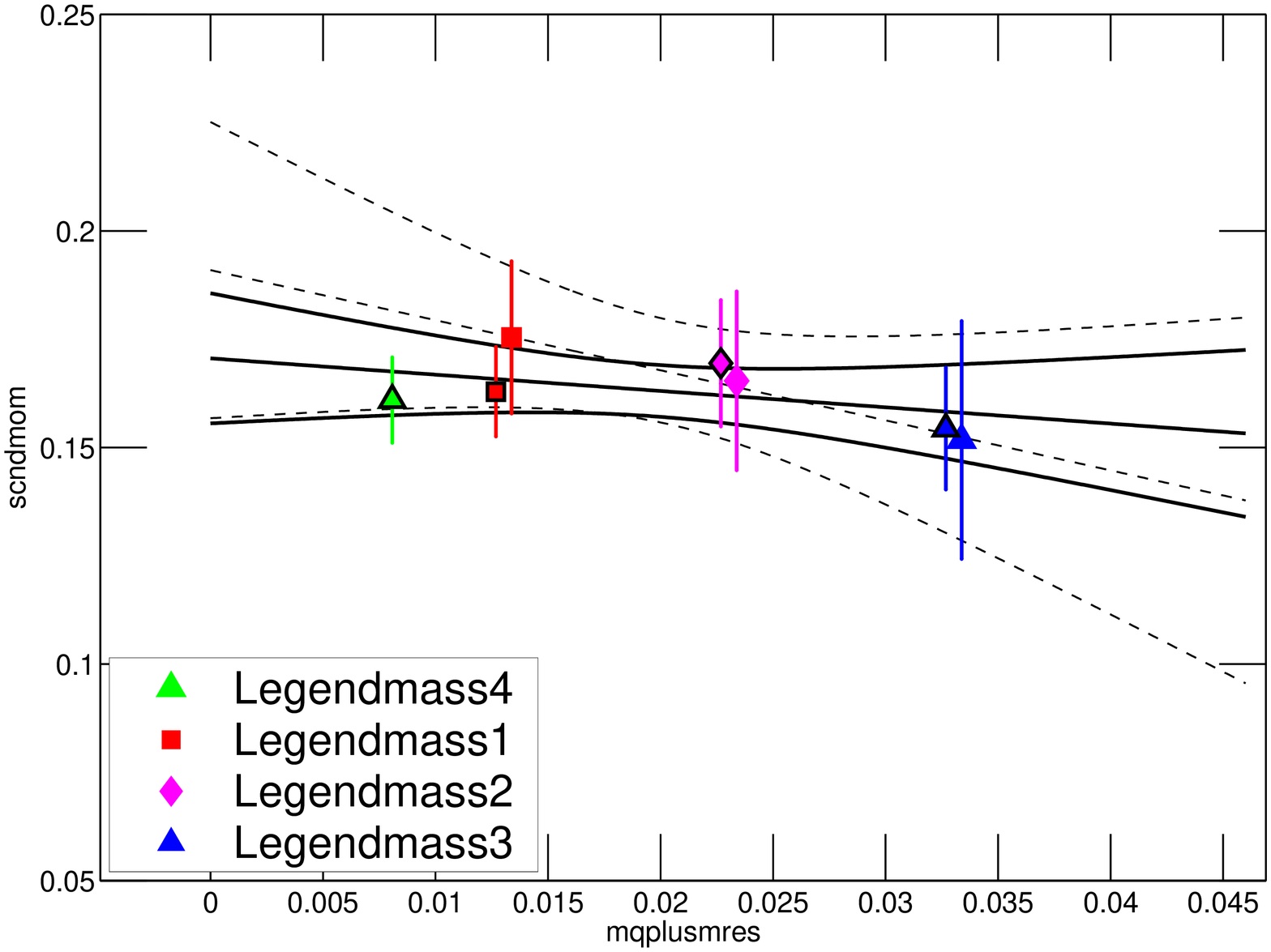}
\end{center}\end{minipage}\hspace{0.039\linewidth}
\begin{minipage}{.48\linewidth}\begin{center}
  \psfrag{mqplusmres}[t][c][1][0]{\footnotesize$am_q+am_{\rm res}$}
  \psfrag{scndmom}[b][c][1][0]{\footnotesize$\xisqav_K^{\rm bare}$}
  \psfrag{Legendmass1}[c][c][0.8][0]{\footnotesize$am_l=0.01$}
  \psfrag{Legendmass2}[c][c][0.8][0]{\footnotesize$am_l=0.02$}
  \psfrag{Legendmass3}[c][c][0.8][0]{\footnotesize$am_l=0.03$}
  \psfrag{Legendmass4}[c][c][0.8][0]{\footnotesize$am_l=0.005$}
  \epsfig{scale=.25,angle=270,file=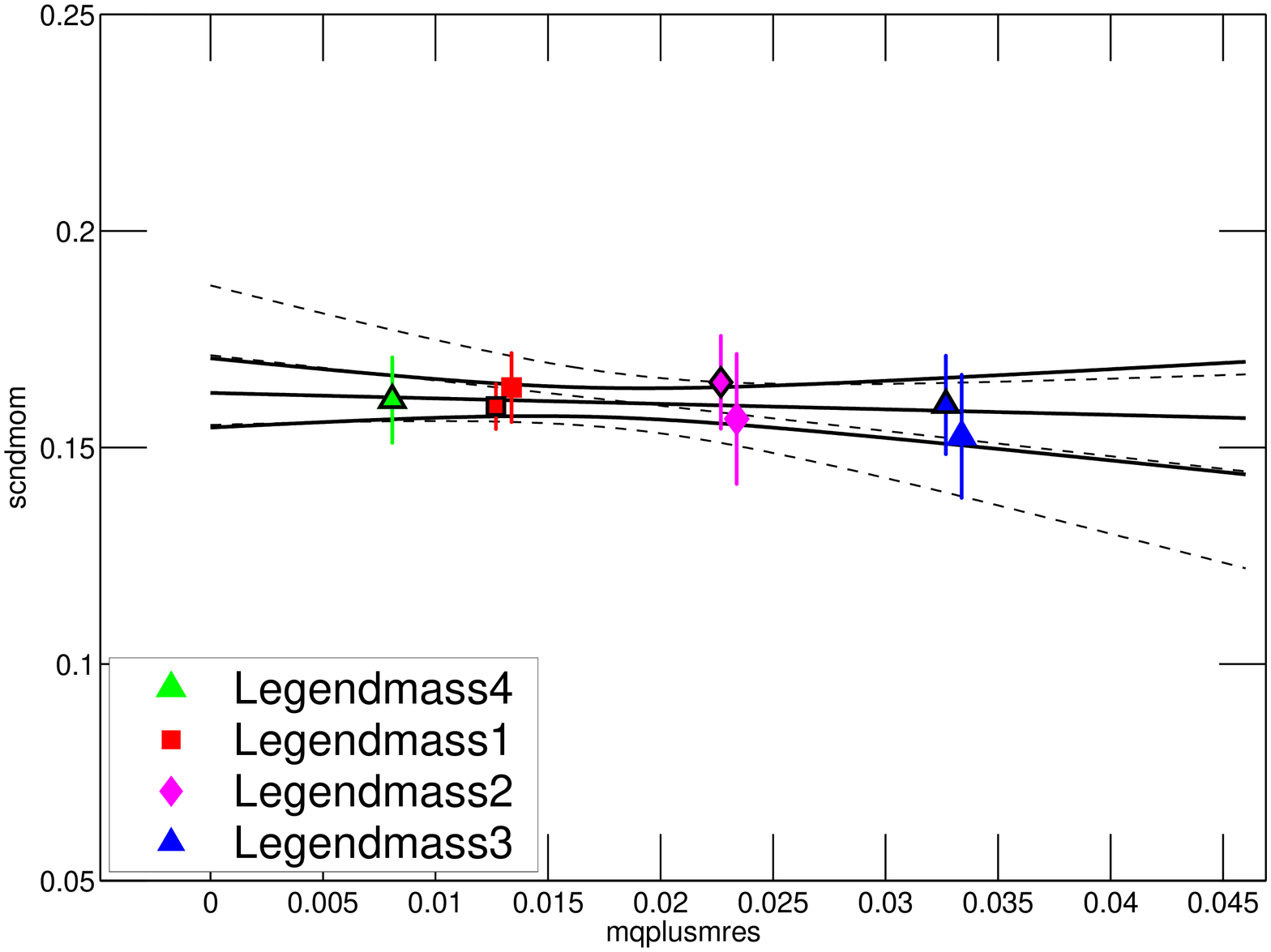}
\end{center}\end{minipage}\caption{Bare values of $\xisqav_\pi$ and
$\xisqav_K$ as a function of the quark
mass.\label{fig:xisqav}}\end{figure}

Chris Sachrajda presented \cite{cts} a calculation of
the first and second moments of distribution amplitudes of the
pion and kaon computed from the following matrix elements
$$
\langle K(q) | \bar{s}(0) \gamma_5 \gamma_{\{\rho}\overset{\leftrightarrow}{D}_{\mu\}}
d(0)|0\rangle = f_K i q_\rho i q_\mu \langle \xi \rangle_K
$$

$$
\langle \pi(q) | \bar{u}(0) \gamma_5 \gamma_{\{\rho}
\overset{\leftrightarrow}{D}_{\mu}
\overset{\leftrightarrow}{D}_{\nu\}}
d(0)|0\rangle = f_\pi i q_\rho i q_\mu i q_\nu \langle \xi^2 \rangle_\pi
.$$
This calculation was performed on our $24^3$ ensembles and follows on from
an earlier work on $16^3$ \cite{Boyle:2006pw,Boyle:2006xq}.
The first moment vanishes in the mass degenerate case and is
non-zero for the kaon, figure~\ref{fig:xiavbare}, but not for
the pion. The second moment
has relatively weak mass dependence for 
both kaon and pion, figure~\ref{fig:xisqav}.
These were renormalised using one-loop lattice perturbation theory and 
we obtain the following preliminary results:

\begin{figure}\begin{center}
\psfrag{ms-minus-mq}[t][c][1][0]{\footnotesize$am_s-am_q$}
  \psfrag{fstmom}[b][t][1][0]{\footnotesize$\xiav_K^{\rm bare}$}
  \psfrag{Legendmass4}[c][c][0.8][0]{\footnotesize$am_{l}=0.005$}
  \psfrag{Legendmass1}[c][c][0.8][0]{\footnotesize$am_{l}=0.01$}
  \psfrag{Legendmass2}[c][c][0.8][0]{\footnotesize$am_{l}=0.02$}
  \psfrag{Legendmass3}[c][c][0.8][0]{\footnotesize$am_{l}=0.03$}
  \epsfig{scale=.25,angle=270,file=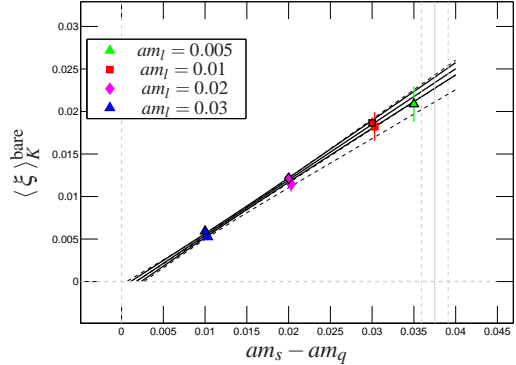}
\caption{Bare values of $\xiav_K$ vs the quark mass. The physical
region $m_sa-m_qa=0.0375(16)$ is marked.
\label{fig:xiavbare}}\end{center}\vspace{-0.2in}\end{figure}

\begin{equation}
\xiav_K^{\overline{\textrm{MS}}}(2\,\textrm{GeV})=0.029(2)\,,\quad
\xisqav^{\overline{\textrm{MS}}}_\pi(2\,\textrm{GeV})=0.28(3)\,,\quad
\xisqav^{\overline{\textrm{MS}}}_K(2\,\textrm{GeV})=0.27(2)\,.
\end{equation}

\section{Vector meson decay constants}
These are defined through
\begin{eqnarray*}
\langle\,0\,|\,\bar{q}_2(0)\gamma^\mu
q_1(0)\,|\,V(p;\lambda)\,\rangle
&=&f_V\,m_V\,\varepsilon_\lambda^\mu\\
\langle\,0\,|\,\bar{q}_2(0)\sigma^{\mu\nu}
q_1(0)\,|\,V(p;\lambda)\,\rangle
&=&if_V^T(\mu)\,\left(\varepsilon_\lambda^\mu p^\nu-
\varepsilon_\lambda^\nu p^\mu\right)\,.
\end{eqnarray*}

The vector meson decay consnant $f_V$ is well constrained
experimentally, but the tensor current coupling is useful
input that lattice gauge theory can provide to sum rules
and other phenomenological applications. Chris Sachrajda
presented a 2+1f DWF calculation of these transverse decay constants,
renormalised with RI-mom NPR using both $16^3$ and $24^3$ volumes.

These results were obtained using only the input strange quark mass of
$0.04$, rather than the 
more physically realistic input quark mass
(i.e. not including $m_{\rm res}$)
of $0.0343$. The ratios $\frac{f_V^T}{f_V}$
display very weak dependence on $m_{l}$ and an estimate absorbed this
change in strange quark mass to the physical point.
As the tensor current is scheme and scale dependent the final results were quoted
at 2 GeV in the $\overline{MS}$ scheme as

\begin{equation}
\frac{f_V^T(2\,\textrm{GeV})}{f_V}=\frac{Z_T(2\,\textrm{GeV}a)}{Z_V}
\,\frac{f_V^{T\,\textrm{bare}}(a)}{f^{\textrm{bare}}_V}=1.11(1)\,
\frac{f_V^{T\,\textrm{bare}}(a)}{f^{\textrm{bare}}_V}\,.\end{equation}
In the $\overline{\textrm{MS}}$ scheme with $\mu=2$\,GeV we
finally obtain:
\begin{equation}\frac{f_\rho^T}{f_\rho}=0.681(20);\quad
\frac{f_{K^\ast}^T}{f_{K^\ast}}=0.712(11);\quad
\frac{f_{\phi}^T}{f_\phi}=0.751(9)\,.\label{eq:finalpreliminary}
\end{equation}

\section{Nucleon mass and structure}

\begin{figure}
\begin{center}
\includegraphics*[width=0.6\textwidth]{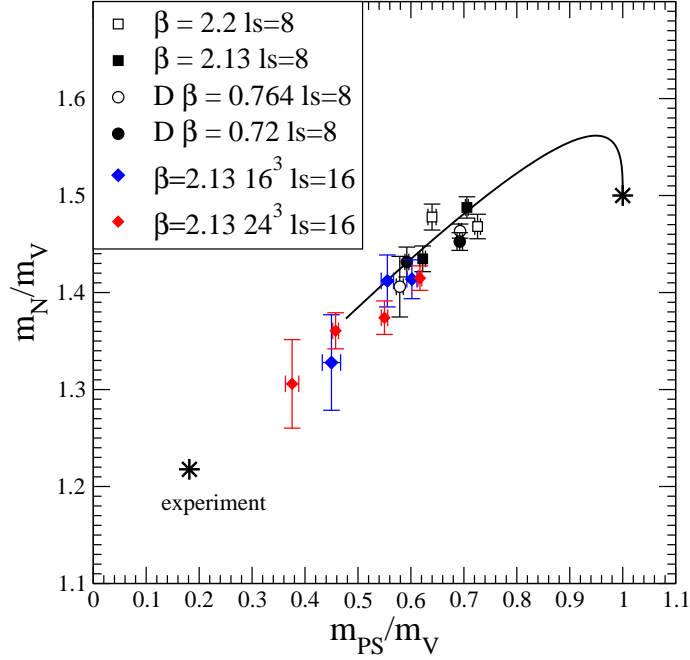}
\end{center}
\caption{Edinburgh plot obtained from various RBC and UKQCD joint ensembles.
The DBW2 gauge action was used with $\beta=0.764$ and $0.72$ while the Iwasaki gauge
action was used with $\beta=2.13$ and $\beta=2.2$. The red data points represent
the nucleon masses for $m_l\in \{0.005,0.01,0.02,0.03\}$ for our $24^3$
ensembles. In the absence of a controlled extrapolation including chiral
non-analyticities, our 2+1f results suggest plausible agreement with experiment
for Nucleon masses, and also suggest reasonable scaling behaviour across several
couplings and gauge actions. 
}
\label{fig:EdinburghPlot}
\end{figure}

Takeshi Yamazaki and Shigemi Ohta have presented results \cite{ohta} for 
isovector form factors and low moments of structure functions of the nucleon,
and related work has also been discussed at other conferences 
\cite{Lin:2007gv,Lin:2006xt}.
These are performed using our four $24^3$ ensembles and the corresponding
Edinburgh plot \cite{BaryonPaper} is displayed in figure~\ref{fig:EdinburghPlot}.

\begin{figure}
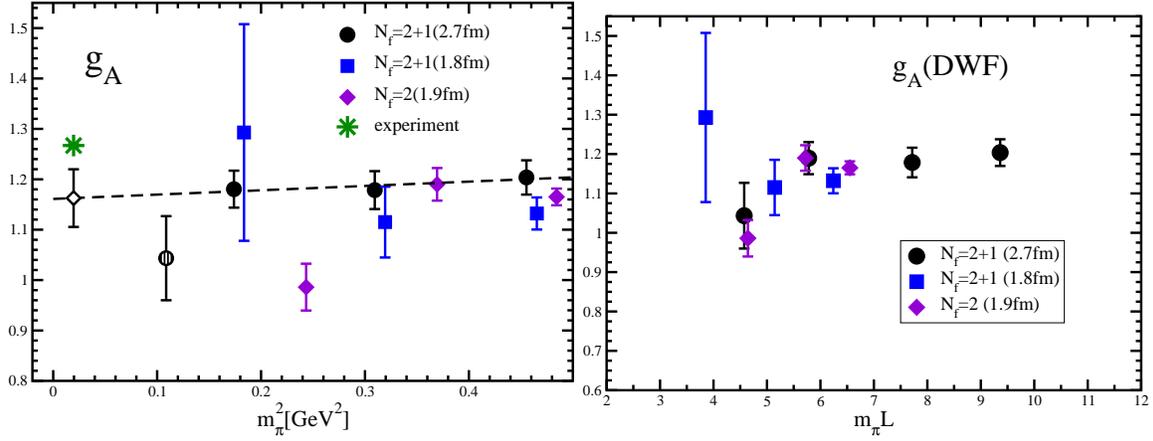

\includegraphics*[width=0.5\textwidth]{mfgagv.eps}
\includegraphics*[width=0.5\textwidth]{mpiL_D.eps}
\caption{
Left panel shows our results for $g_A$ as a function of $m_\pi$ for several ensembles.
Potential finite volume effects explain differences between these, and a deviation from
experiment. The right panel displays the same data as a function of $m_\pi L$ and the
scaling of the deviant points with volume becomes apparent. The rather poorly determined
blue result for our $16^3$ ensembles needs more effort to establish whether it confirms this picture.
}
\label{fig:gAmpi}
\label{fig:gAmpiL}
\end{figure}

Nucleon three point functions have been calculated using a source-sink
time separation of 12.
For our $(2.7 {\rm fm})^3$ simulation we find, figure~\ref{fig:gAmpi},
that the axial charge $g_A$ appears flat except at
our lightest datapoint, which is around 15\% lower. Similar behavour was seen
by RBC, at heavier masses on a $(1.9 {\rm fm})^3$ 2-flavour DWF simulation. We believe
this is a finite volume effect with the mass threshold determined by the volume.
Our $m_u = 0.01$ data point on our $16^3$ ensemble does not display this effect,
but carries very large statistical errors.
Figure~\ref{fig:gAmpiL} displays our results against $m_\pi L$, and is
suggestive of mass dependent finite volume effects which appear
to scale with $m_\pi L$ and appear for $m_\pi L \lesssim 6$.
We observe similar behaviour if we plot the results with
Wilson fermions by LHPC/SESAM and QCDSF in this fashion 
\cite{ohta,Dolgov:2002zm,Khan:2006de}.
Improved statistics for the $16^3$ 2+1f measurements is important to clarify
the one (statistically questionable) exception to this picture.

The lowest $24^3$ data point is omitted from an extrapolation, and $g_A=1.16(6)$
obtained at the physical pion mass.
Results were also presented for the vector, axial, induced tensor
and induced pseudoscalar form factors, some associated couplings
(such as $g_{\pi N N}$ and the induced pseudoscalar coupling $g_P$)
and corresponding mean squared radii. The momentum fraction, 
helicity fraction, transversity 
and twist-3 $d_1$ structure function moments were calculated.


\section{Conclusions}

RBC and UKQCD have exploited the PPARC, Riken, and SciDAC QCDOC machines in Edinburgh and Brookhaven 
to simulate dynamical domain wall fermions with realistic sea quark content. The analysis of the
first lattice spacing $a^{-1}=1.73$ GeV on a $(2.7 {\rm fm})^3$ volume is 
well advanced with a broad and rich physics programme presented
at this conference. The programme will continue to analyse two ensembles on a finer lattice spacing 
that are currently being generated.
The physics parameters are very competitive despite the cost of the additional
fifth dimension, with sea pion masses down to $330$ MeV and valence pions down to $240$ MeV.
Partially quenched chiral perturbation theory is exploited in our analysis programme
and enabled many more data points to be measured within the SU(2)$\times$SU(2) chiral regime.
We have obtained results for the low energy constants of the chiral effective lagrangian, 
quark masses, $B_K$, and $V_{us}$ from both $f_K/f_\pi$ and from Kl3. 
We find that very large lattice volumes $m_\pi L \ge 6 $ may be required for non-spectral
nucleon physics based on suspected finite volume effects in the nucleon axial charge.

\section{Acknowledgements}

I wish to thank my colleagues in the RBC and UKQCD collaborations whose
work it has been my priviledge to review:
Conrado Albertus,
Chris Allton,
Dave Antonio,
Yasumichi Aoki,
Christopher Aubin,
Tom Blum,
Ken Bowler,
Michael Cheng,
Norman Christ,
Michael Clark,
Saul Cohen,
Paul Cooney,
Chris Dawson,
Luigi del Debbio,
Michael Donellan,
Mike Endres,
Jonathan Flynn,
Alistair Hart,
Koichi Hashimoto,
Tomomi Ishikawa,
Taku Izubuchi,
Xiao-Yong Jin,
Chulwoo Jung,
Andreas J\"uttner,
Tony Kennedy,
Richard Kenway,
Changhoan Kim,
Min Li,
Sam Li,
Adam Lichtl,
Hugo H M C Pedroso de Lima,
Huey Wen Lin,
Meifeng Lin,
Oleg Loktik,
Robert Mawhinney,
Chris Maynard,
Jun Ichi Noaki,
Shigemi Ohta,
Brian Pendleton,
Chris Sachrajda,
Shoichi Sasaki,
Enno Scholz,
Amarjit Soni,
Aurora Trivini,
Robert Tweedie,
Jan Wennekers,
Azusa Yamaguchi,
Takeshi Yamazaki, and
James Zanotti.
We thank the QCDOC design team for developing with us the QCDOC
machine and its software. This development and the computers used in 
this calculation were funded by the U.S.\ DOE grant DE-FG02-92ER40699, 
PPARC JIF grant PPA/J/S/1998/00756 and by RIKEN. This work was supported 
by DOE grant DE-FG02-92ER40699 and PPARC grants PPA/G/O/2002/00465 and 
PP/D000238/1. We thank the University of Edinburgh, PPARC, RIKEN, BNL 
and the U.S.\ DOE for providing the QCDOC facilities.

\end{document}